\documentclass{JHEP3}
\usepackage{amssymb}
\usepackage{graphicx}
\usepackage{epsfig}
\usepackage{axodraw}
\usepackage{citesort}
\newcommand{\bmat}{\left(\begin{array}}
\newcommand{\emat}{\end{array}\right)}
\newcommand{\be}{\begin{equation}}
\newcommand{\ee}{\end{equation}}
\newcommand{\bea}{\begin{eqnarray}}
\newcommand{\eea}{\end{eqnarray}}
\def\ie{{\it i.e.}}
\def\lsim{\raise0.3ex\hbox{$\;<$\kern-0.75em\raise-1.1ex\hbox{$\sim\;$}}}
\def\gsim{\raise0.3ex\hbox{$\;>$\kern-0.75em\raise-1.1ex\hbox{$\sim\;$}}}

\date{\today}
\title{Higgs and $Z^{\prime }$ Phenomenology in $B-L$ extension of the Standard Model at
LHC}

\author{W. Emam and S. Khalil\\
Center for Theoretical Physics at the
British University in Egypt, Sherouk City, Cairo 11837, Egypt.\\
Faculty of Science, Ain Shams University, Cairo 11566, Egypt.}

\abstract{The phenomenology of the low scale $U(1)_{B-L}$
extension of the standard model and its implications at LHC is
presented. In this model, an extra gauge boson corresponding to
$B-L$ gauge symmetry and an extra SM singlet scalar (heavy Higgs)
are predicted. We show a detailed analysis of both heavy and light
Higgses decay and production in addition to the possible decay
channels of the new gauge boson. We find that the cross sections
of the SM-like Higgs production are reduced by $\sim 20\%-30\%$,
while its decay branching ratios remain intact. The extra Higgs
has relatively small cross sections and the branching ratios of
$Z^{\prime} \to l^+ l^-$ are of order $\sim 20\%$ compared to
$\sim 3\%$ of the SM resuls. Hence, the search for $Z^{\prime}$ is
accessible via a clean dilepton signal at LHC. }


\preprint{}
\begin{document}
\section{Introduction}

The Standard Model (SM) of elementary particles has been regarded only as a
low energy effective theory of the yet-more-fundamental theory. Several
attempts have been proposed to extend the gauge symmetry of the SM via one
or more $U(1)$ gauge symmetries beyond the hypercharge gauge symmetry, $%
U(1)_{Y}$~\cite{Mohapatra:1980qe,Khalil:2006yi,Cerdeno:2006ha}.
The evidence for non-vanishing neutrino masses, based on the
apparent observation of neutrino oscillation, strongly encourages
this type of extensions. In this class of
models~\cite{Mohapatra:1980qe,Khalil:2006yi}, three SM singlet
fermions arise quite naturally due to the anomaly cancellation
conditions. These three particles are accounted for right handed
neutrinos, and hence a natural explanation for the seesaw
mechanism is obtained.

A low scale $B-L$ symmetry breaking, based on the gauge group
$G_{B-L}\equiv SU(3)_{C}\times SU(2)_{L}\times U(1)_{Y}\times
U(1)_{B-L}$, has been considered recently \cite{Khalil:2006yi}. It
was shown that this model can account for the current experimental
results of the light neutrino masses and their large mixing.
Therefore, it can be considered as one of the strong candidates
for minimal extensions of the SM. In addition, one extra neutral
gauge boson corresponding to $B-L$ gauge symmetry and an extra SM
singlet scalar (extra Higgs) are predicted. In fact, the SM Higgs
sector can be generally extended by adding extra singlet scalars
without enlarging its gauge symmetry
group~\cite{O'Connell:2006wi,Datta:1997fx}. In
Ref.\cite{Khalil:2006yi}, it has been emphasized that these new
particles may have significant impact on the SM phenomenology,
hence lead to interesting signatures at Large Hadron Collider
(LHC).

The aim of this paper is to provide a comprehensive analysis for
the phenomenology of such TeV scale extension of the SM, and its
potential discovery at the LHC. The production cross sections and
the decay branching ratios of the SM like Higgs, $H$, and the
extra Higgs boson $H^{\prime}$ are analyzed. We also consider the
decay branching ratios of the extra gauge boson, $Z^{\prime }$.

We show that the cross sections of the Higgs production are
reduced by $\sim 20\%-30\%$ in the interesting mass range of $\sim
120 -250$ GeV relative to the SM predictions. However, its decay
branching ratios remain intact. In addition, we find that the
extra Higgs ($\sim$ TeV) is accessible at LHC, although it has
relatively small cross sections. We also examine the availability
of the decay channel $H^{\prime } \to HH$, which happens to have
very small partial decay width. Concerning the $Z^{\prime }$ gauge
boson, the branching ratios of $Z^{\prime } \to l^+ l^-$ are found
to be of order $\sim 20\%$ compared to $\sim 3\%$ of the SM
$BR(Z\to l^+ l^-)$.

This paper is organized as follows. In section 2 we review the
Higgs mechanism and symmetry breaking within the minimal $B-L$
extension of the SM. We also discuss the mixing between the
SM-like Higgs and the extra Higgs boson. Section 3 is devoted for
the phenomenology of the two Higgs particles. The production cross
sections and decay branching ratios of these Higgs particles at
LHC are presented. In section 4 we study the decay of the extra
gauge boson $Z^{\prime}$. In section 5 we briefly discuss the
scenario of {\it very light Higgs}. Finally we give our concluding
remarks in section 6.

\section{$B-L$ extension of the SM}

\subsection{Symmetry breaking}

The fermionic and kinetic sectors of the Lagrangian\ in the case of $B-L$
extension are given by%
\begin{eqnarray}
\mathcal{L}_{B-L} &=&i~\bar{l}D_{\mu }\gamma ^{\mu }l+i~\bar{e}_{R}D_{\mu
}\gamma ^{\mu }e_{R}+i~\bar{\nu}_{R}D_{\mu }\gamma ^{\mu }\nu _{R}  \nonumber
\\
&&-\frac{1}{4}W_{\mu \nu }W^{\mu \nu }-\frac{1}{4}B_{\mu \nu }B^{\mu \nu }-%
\frac{1}{4}C_{\mu \nu }C^{\mu \nu }.
\end{eqnarray}%
The covariant derivative $D_{\mu }$ is different from the SM one by the term
$ig^{\prime \prime }Y_{B-L}C_{\mu }$, where $g^{\prime \prime }$ is the $%
U(1)_{B-L}$ gauge coupling constant, $Y_{B-L}$ is the $B-L$ charge, and $%
C_{\mu \nu }=\partial _{\mu }C_{\nu }-\partial _{\nu }C_{\mu }$ is
the field strength of the $U(1)_{B-L}$. The $Y_{B-L}$ for fermions
and Higgs are given in Table~\ref{tab:b-l}.
\begin{table}[thbp]
\centering {\large
\begin{tabular}{ccccccc}
\hline\hline particle & $l$ & $e_{R}$ & $\nu _{R}$ & $q$ & $\phi $
& $\chi $ \\ \hline $Y_{B-L}$ & $-1$ & $-1$ & $-1$ & $1/3$ & $0$ &
$2$ \\ \hline\hline
\end{tabular}%
} \caption{$B-L$ quantum numbers for fermions and Higgs particles}
\label{tab:b-l}
\end{table}

The Higgs and Yukawa sectors of the Lagrangian are given by
\begin{eqnarray}
\mathcal{L}_{B-L} &=&(D^{\mu }\phi )(D_{\mu }\phi )+(D^{\mu }\chi )(D_{\mu
}\chi )-V(\phi ,\chi )  \nonumber \\
&&-\Big(\lambda _{e}\bar{l}\phi e_{R}+\lambda _{\nu }\bar{l}\tilde{\phi}{\nu
}_{R}+\frac{1}{2}\lambda _{\nu _{R}}\bar{\nu ^{c}}_{R}\chi \nu _{R} +h.c.%
\Big).
\end{eqnarray}%
Here, $\lambda _{e}$, $\lambda _{\nu }$ and $\lambda _{\nu _{R}}$ refer to $%
3\times 3$ Yakawa matrices. The interaction terms $\lambda _{\nu }l\tilde{%
\phi}\nu _{R}$ and $\lambda _{\nu _{R}}\bar{\nu ^{c}}_{R}\chi \nu _{R}$\
give rise to a Dirac neutrino mass term: $m_{D}\simeq \lambda _{\nu }v$ and
a Majorana mass term: $M_{R}=\lambda _{\nu _{R}}v^{\prime }$, respectively.
The $U(1)_{B-L}$ and $SU(2)_{L}\times U(1)_{Y}$ gauge symmetries can be
spontaneously broken by a SM singlet complex scaler field $\chi $ and a
complex $SU(2)$ doublet of scalar fields $\phi $, respectively. We consider
the most general Higgs potential invariant under these symmetries, which is
given by%
\begin{eqnarray}
V(\phi ,\chi ) &=&m_{1}^{2}\phi ^{\dagger }\phi +m_{2}^{2}\chi ^{\dagger
}\chi +\lambda _{1}(\phi ^{\dagger }\phi )^{2}+\lambda _{2}(\chi ^{\dagger
}\chi )^{2}  \nonumber \\
&&+\lambda _{3}(\chi ^{\dagger }\chi )(\phi ^{\dagger }\phi ),
\end{eqnarray}%
where $\lambda _{3}>-2\sqrt{\lambda _{1}\lambda _{2}}$ and
$\lambda _{1},\lambda _{2}\geq 0$, so that the potential is
bounded from below. For non-vanishing vacuum expectation values
(vev's), we require $\lambda _{3}^{2}<4\lambda _{1}\lambda _{2}$ ,
$m_{1}^{2}<0$ and $m_{2}^{2}<0$. The vev's, $|\langle \phi \rangle
|=v/\sqrt{2}$ and $|\langle \chi \rangle |=v^{\prime }/\sqrt{2}$,
are then given by
\[
v^{2}=\frac{4\lambda _{2}m_{1}^{2}-2\lambda _{3}m_{2}^{2}}{\lambda
_{3}^{2}-4\lambda _{1}\lambda _{2}},~\ \ \ \ ~~v^{\prime 2}=\frac{%
-2(m_{1}^{2}+\lambda _{1}v^{2})}{\lambda _{3}}.
\]%
Depending on the value of the $\lambda _{3}$ coupling, one can have $%
v^{\prime }\gg v$ or $v^{\prime }\approx v$. Therefore, the symmetry
breaking scales, $v$ and $v^{\prime }$, can be responsible for two different
symmetry breaking scenarios. In our analysis we take $v=246$ GeV and
constrain the other scale, $v^{\prime }$, by the lower bounds imposed on the
mass of the extra neutral gauge boson.

After the $B-L$ gauge symmetry breaking, the gauge field $C_{\mu}$
(will be called  $Z^{\prime}$ in the rest of the paper) acquires
the following mass:
\begin{equation}
m_{Z^{\prime }}^{2}=4g^{\prime \prime }v^{\prime 2}.
\end{equation}%
The experimental search for $Z^{\prime }$ at CDF experiment leads to $%
m_{Z^{\prime }} \gsim O(600)$ GeV. However, the strongest limit
comes from LEP II \cite{Carena:2004xs}:
\begin{equation}
m_{Z^{\prime }}/g^{\prime \prime }>6{TeV}.
\end{equation}%
This implies that $v^{\prime }\gsim O($TeV$)$. Moreover, if the coupling $%
g^{\prime \prime }$ is $<O(1)$, one can still obtain $m_{Z^{\prime
}} \gsim O(600) $ GeV.

\subsection{Higgs sector}

In addition to the SM complex $SU(2)_{L}$ doublet, another complex scalar
singlet arise in this class of models. Out of these six scalar degrees of
freedom, only two physical degrees of freedom,\ ($\phi ,\chi $), remain
after the $B-L$ and electroweak symmetries are broken. The other four
degrees of freedom are eaten by $Z^{\prime }$, $Z$ and $W^{\pm }$ bosons.

The mixing between the two Higgs scalar fields is controlled by the coupling
$\lambda _{3}$. In fact, one finds that for positive $\lambda _{3}$ , the $%
B-L$ symmetry breaking scale, $v^{\prime }$, becomes much higher than the
electroweak symmetry breaking scale, $v$. In this case, the SM \ singlet
Higgs, $\phi $, and the SM like Higgs, $\chi $, are decoupled and their
masses are given by
\begin{equation}
M_{\phi }=\sqrt{2\lambda _{1}}v ,~\ \ \ \ ~ M_{\chi
}=\sqrt{2\lambda _{2}}v^{\prime }.
\end{equation}%
For negative $\lambda _{3}$, however, the $B-L$ breaking scale is at the
same order of the the electroweak breaking scale. In this scenario, a
significant mixing between the two Higgs scalars exists and can affect the
SM phenomenology. This mixing can be represented by the following mass
matrix for $\phi $ and $\chi $:
\begin{equation}
\frac{1}{2}M^{2}(\phi ,\chi )=\left(
\begin{array}{cc}
\lambda _{1}v^{2} & \frac{\lambda _{3}}{2}vv^{\prime } \\
\frac{\lambda _{3}}{2}vv^{\prime } & \lambda _{2}v^{\prime 2}%
\end{array}%
\right) .
\end{equation}%
Therefore, the mass eigenstates fields $H$ and $H^{\prime }$ are given by
\begin{equation}
\left(
\begin{array}{c}
H \\
H^{\prime }%
\end{array}%
\right) =\left(
\begin{array}{cc}
\cos \theta  & -\sin \theta  \\
\sin \theta  & \cos \theta
\end{array}%
\right) \left(
\begin{array}{c}
\phi  \\
\chi
\end{array}%
\right) ,
\end{equation}%
where the mixing angle $\theta $ is defined by
\begin{equation}
\tan 2\theta =\frac{|\lambda _{3}|vv^{\prime }}{\lambda _{1}v^{2}-\lambda
_{2}v^{\prime 2}}.
\end{equation}%
The masses of $H$ and $H^{\prime }$ are given by
\begin{equation}
m_{H,H^{\prime }}^{2}=\lambda _{1}v^{2}+\lambda _{2}v^{\prime 2}\mp \sqrt{%
(\lambda _{1}v^{2}-\lambda _{2}v^{\prime 2})^{2}+\lambda
_{3}^{2}v^{2}v^{\prime 2}}.
\end{equation}%
We call $H$ and $H^{\prime }$ as light and heavy Higgs bosons,
respectively. In our analysis we consider a maximum mixing between
the two Higgs bosons by taking $\left\vert \lambda _{3}\right\vert
\simeq \lambda _{1}^{\max }\lambda _{2}^{\max }$, where $\lambda
_{1}^{\max }$ and $\lambda _{2}^{\max }$ are given by
\begin{eqnarray}
\lambda _{1}^{\max } &=&\frac{m_{H}^{2}+m_{H^{\prime }}^{2}-\sqrt{%
4m_{H}^{2}m_{H^{\prime }}^{2}+1}+1}{4v^{2}},  \nonumber \\
\lambda _{2}^{\max } &=&\frac{m_{H}^{2}+m_{H^{\prime }}^{2}+\sqrt{%
4m_{H}^{2}m_{H^{\prime }}^{2}+1}-1}{4v^{\prime 2}},
\end{eqnarray}%
and the maximum mixing angle is then given by
\begin{equation}
\tan 2\theta =\frac{\lambda _{1}^{\max }\lambda _{2}^{\max }vv^{\prime }}{%
\lambda _{1}^{\max }v^{2}-\lambda _{2}^{\max }v^{\prime 2}}.
\end{equation}

\begin{figure}[t]
\begin{center}
\epsfig{file=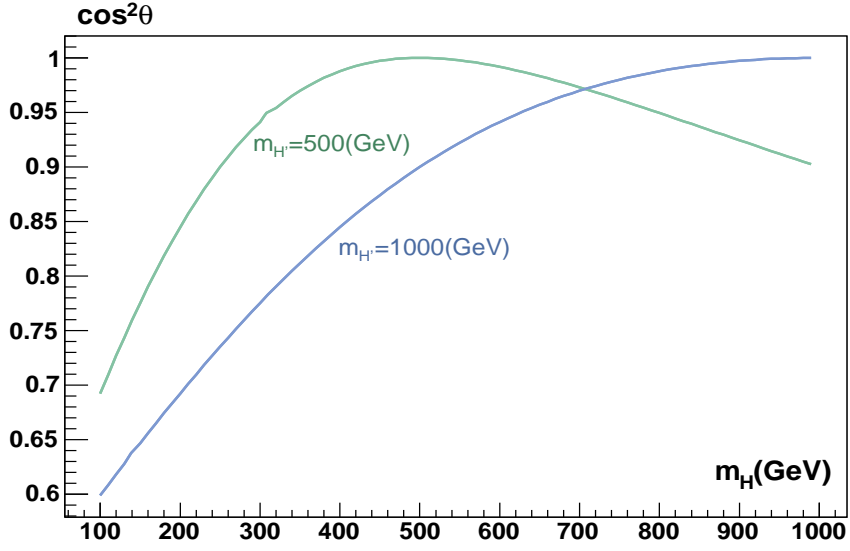, width=13cm, height=8cm, angle=0}
\end{center}
\par
\vskip-0.5cm \caption{$H-H^{\prime}$ mixing angle as function of
$m_H$ for $m_H^{\prime}=500$ GeV and $1$ TeV.} \label{fig:theta}
\end{figure}

By considering the maximum mixing and fixing $v=246$ GeV and
$v'=1$ TeV, we have reduced the number of free parameters of this
model into just two, namely $m_{H}$ and $m_{H^{\prime}}$. In
Figure \ref{fig:theta}, we present the maximum mixing as a
function of the light Higgs mass, $m_{H}$ for $m_{H^{\prime}}=500$
GeV and $1$ TeV.

Due to the mixing between the two Higgs bosons, the usual couplings among the SM-like Higgs, $%
H $, and the SM fermions and gauge bosons are modified. In
addition, there are new couplings among the extra Higgs,
$H^{\prime} $, and the SM particles:
\begin{eqnarray}
g_{Hff}=i\frac{m_{f}}{v}\cos \theta , &~~&g_{H^{\prime }ff}=i\frac{m_{f}}{v}%
\sin \theta ,  \nonumber \\
g_{HVV}=-2i\frac{m_{V}^{2}}{v}\cos \theta , &~~&g_{H^{\prime }VV}=-2i\frac{%
m_{V}^{2}}{v}\sin \theta ,  \nonumber \\
g_{HZ^{\prime }Z^{\prime }}=2i\frac{m_{C}^{2}}{v^{\prime }}\sin
\theta , &~~&g_{H^{\prime }Z^{\prime }Z^{\prime
}}=-2i\frac{m_{C}^{2}}{v^{\prime }}\cos \theta ,
\nonumber \\
g_{H\nu _{R}\nu _{R}}=-i\frac{m_{\nu _{R}}}{v^{\prime }}\sin \theta ,
&~~&g_{H^{\prime }\nu _{R}\nu _{R}}=i\frac{m_{\nu _{R}}}{v^{\prime }}\cos
\theta .
\end{eqnarray}

The Higgs self couplings are give by
\begin{eqnarray}
g_{H^{3}} &=&6i(\lambda _{1}v\cos ^{3}\theta -\frac{\lambda _{3}}{2}%
v^{\prime} \cos^2\theta \sin \theta ),  \nonumber \\
g_{H^{\prime 3}} &=&6i(\lambda _{2}v^{\prime} \cos^3\theta +\frac{\lambda _{3}}{2}%
v\cos ^{2}\theta \sin \theta ),  \nonumber \\
g_{H^{4}} &=&6i\lambda _{1}\cos^{4}\theta ,  \nonumber \\
g_{H^{\prime 4}} &=&6i\lambda _{2}\cos^{4}\theta ,  \nonumber \\
g_{HH^{\prime 2}} &=&2i(\frac{\lambda_{3}}{2}v\cos^{3}\theta
+\lambda_{3}v^{\prime} \cos^2\theta \sin \theta
-3\lambda_{2}v^{\prime}\cos^2\theta \sin
\theta),  \nonumber \\
g_{H^{2}H^{\prime}}
&=&2i(\frac{\lambda_{3}}{2}v^{\prime}\cos^3\theta
-\lambda_{3}v\cos ^{2}\theta \sin \theta +3\lambda _{1}v\cos
^{2}\theta \sin \theta
),  \nonumber \\
g_{H^{2}H^{\prime 2}} &=&i\lambda _{3}\cos^{4}\theta .
\label{Hcouplings}
\end{eqnarray}

These new couplings lead to a different Higgs phenomenology from
the well known one, predicted by the SM. The detailed analysis of
Higgs bosons in this class of models and their phenomenological
implications, like their productions and decays at the LHC, will
be discussed in the next section.

\section{Higgs Production and Decay at Hadron Colliders}

\subsection{Higgs Production}

At the LHC, two 7-TeV proton beams with a center-of-mass energy of
14 TeV and a luminosity of $10^{34}$cm$^{-2}s^{-1}$ will collide
with each other. The machine is expected to start running early
2008. The detection of the SM Higgs boson is the primary goal of
the LHC project.

At hadron colliders, the two Higgs bosons couple mainly to the
heavy particles: the massive gauge bosons $Z^{\prime }$, $Z$ and
$W^{\pm }$ and the heavy quarks $t$, $b$. The main production
mechanisms for Higgs particles can be classified into four
groups~\cite{Djouadi:2005gi}: the gluon--gluon fusion
mechanism\cite{Georgi:1977gs}, the associated Higgs production
with heavy
top or bottom quarks\cite{Raitio:1978pt}, the associated production with $%
W/Z/Z^{\prime }$ bosons\cite{Glashow:1978ab}, and the weak vector
boson fusion processes\cite{Dicus:1985zg}:
\begin{eqnarray}
gg &\to& H \label{ggH}\\
gg, q \bar{q} &\to& Q \bar{Q} + H, \label{HQQ}\\
q \bar{q} &\to&  V + H \label{VH}\\
q q & \to & V^* V^* \to qq + H \label{qqH}.
\end{eqnarray}

\begin{figure}[t]
\resizebox{9.5in}{!}{\includegraphics*[0.5in,7.0in][12in,10in]{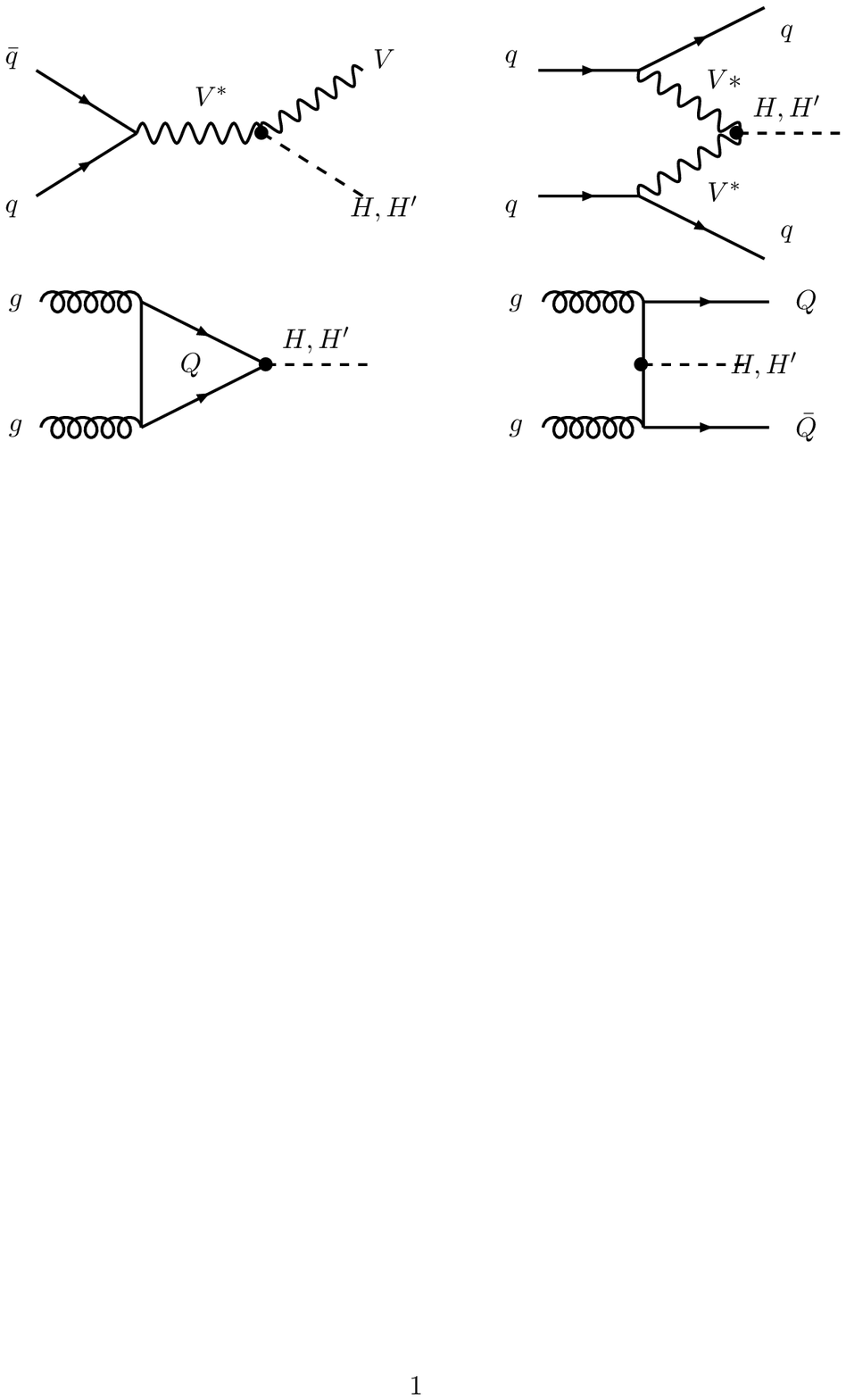}}
\vspace*{-7mm} \caption{The dominant Higgs boson production
mechanisms in hadronic collisions.} \label{fig:Fyn1}
\end{figure}
\vspace*{4mm}
The Feynman diagrams of these processes are displayed in Figure~\ref%
{fig:Fyn1}. The cross sections of the Higgs production in these
four mechanisms are directly proportional to the the Higgs
couplings with the associated particles.

In case of the gluon--gluon fusion mechanism the Higgs production
is mediated by triangular loops of heavy quarks. Thus, the cross
section of this process is proportional to the Higgs coupling with
the heavy quark mass.  In case of $B-L$ extension of the SM, the
production cross sections for the light Higgs, $H$, and the heavy
Higgs, $H^{\prime }$, can be approximated as
\begin{eqnarray}
\sigma _{H} &\propto& \alpha^2_s \left(\frac{m_{Q}^{2}}{v^2}\cos
^{2}\theta
\right)~ \times ~\left(\frac{m_Q^2}{m_H^2}\right),\label{sigmaH}\\
\sigma _{H^{\prime }}&\propto& \alpha^2_s
\left(\frac{m_{Q}^{2}}{v^2}\sin ^{2}\theta\right)~ \times ~
\left(\frac{m_Q^2}{m^2_{H^{\prime}}}\right),\label{sigmaH'}
\end{eqnarray}
where the first bracket is due to the coupling $QQH(H^{\prime})$,
while the second bracket corresponds to an approximated loop
factor. As can be seen from Equations~\ref{sigmaH} and
\ref{sigmaH'}, the cross section of the light Higgs production is
reduced respect to the SM one by the factor of $\cos ^{2}\theta $.
On the other hand, the heavy Higgs production is suppressed by two
factors: the small $\sin\theta$, and the large $m_{H^{\prime}}$.
Therefore, the the heavy Higgs production is typically less than
that of the light Higgs by two orders of magnitudes, \ie,
\be%
\frac{\sigma_{H^{\prime}}}{\sigma_H} \simeq
\frac{\sin\theta^2}{\cos\theta^2} ~
\frac{m_H^2}{m^2_{H^{\prime}}} \simeq {\cal O}(10^{-2}).%
\label{ratio} \ee

\begin{figure}[t]
\hspace{-1cm}\resizebox{9.5in}{!}{\includegraphics*[0.5in,8.5in][12in,10in]{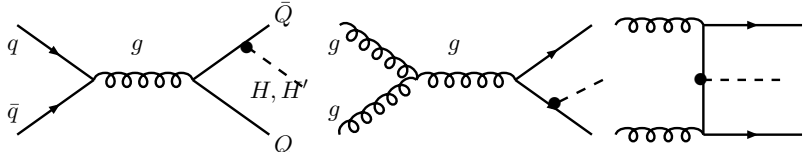}}
\vspace*{-6mm} \caption{Feynman diagrams for Higgs production in
association with heavy
quarks in hadronic collisions, $pp\rightarrow q\bar{q},gg\rightarrow Q\bar{Q}%
H$, at LO.}
\label{fig:Fyn2}
\end{figure}

Now, we consider the mechanism of Higgs production in association
with heavy quark pairs, Equation~ \ref{HQQ}. In addition to the
Feynman diagram shown in Figure \ref{fig:Fyn1}, a set of other
diagrams that also contribute to this process is given in
Figure~\ref{fig:Fyn2}.
%
Note that although this process shares the same coupling with the
gluon-gluon fusion process, the leading order expression of its
cross section indicates that it is less by one order of magnitude,
for $m_{H(H^{\prime})} < 1$ TeV. Furthermore, the typical ratio of
$\sigma(gg\to H^{\prime}Q\bar{Q})$ to $\sigma(gg\to H Q\bar{Q})$
is of order $(\sin\theta/\cos\theta)^2 \simeq {\cal O}(0.1)$.

Finally, we study the Higgs production in association with
$W/Z/Z^{\prime }$ bosons and in the weak vector boson fusion
processes, Equations~ \ref{VH} and \ref{qqH} respectively. In
$B-L$ extension of the SM, the cross sections of these channels
are proportional to the mass of the gauge boson and the mixing
angle $\theta$ of the two Higgs bosons:
\begin{eqnarray}
V\equiv W/Z: \sigma _{H}&\propto& \frac{m_{V}^{4}}{v^2}\cos
^{2}\theta \times \frac{g^2}{m_V^2} \times \rm{Loop~ function} ,~\\
\sigma _{H^{\prime}} &\propto& \frac{m_{V}^{4}}{v^2}\sin
^{2}\theta \times \frac{g^2}{m_V^2} \times \rm{Loop~ function}.
\end{eqnarray}

In case of $V\equiv Z^{\prime }$, The production is enhanced by the $%
HZ^{\prime }Z^{\prime }$ coupling arising with $m_{Z^{\prime }}$.
However, it is suppressed by a large value of $v^{\prime }$ and
the mass of the virtual gauge boson(s), $m_{Z^{\prime }}$:
\begin{eqnarray}
V\equiv Z^{\prime }&:&\sigma _{H}\propto \frac{m_{Z^{\prime }}^{4}}{%
v^{\prime^2}}\sin ^{2}\theta \times \frac{(g^{\prime \prime }Y_{B-L}^{Q})^{2}%
}{m_{Z^{\prime }}^{2}} \times \rm{Loop~ function},  \\
&&\sigma _{H^{\prime }}\propto \frac{m_{Z^{\prime
}}^{4}}{v^{\prime^2}}\cos ^{2}\theta \times \frac{(g^{\prime
\prime }Y_{B-L}^{Q})^{2}}{m_{Z^{\prime }}^{2}} \times \rm{Loop~
function}.
\end{eqnarray}

From these equations, one can observe that the relative ratio
between the light Higgs production associated with $W/Z$  and
$Z^{\prime}$ gauge bosons is given by
$\sigma_H(W/Z)/\sigma_H(Z^{\prime}) \sim \cos^2\theta/\sin^2\theta
\times g^{''^2}/g^2(g^{'^2})$. Therefore, $\sigma_H(W/Z)$ can be
larger than $\sigma_H(Z^{\prime})$ by one order of magnitude at
most. In contrary, the situation is reversed for the heavy Higgs
production and one finds that $\sigma_H^{\prime}(Z^{\prime}) >
\sigma_H^{\prime}(W/Z)$, which confirms our earlier discussion.

\begin{figure}[t]
\begin{center}
\epsfig{file=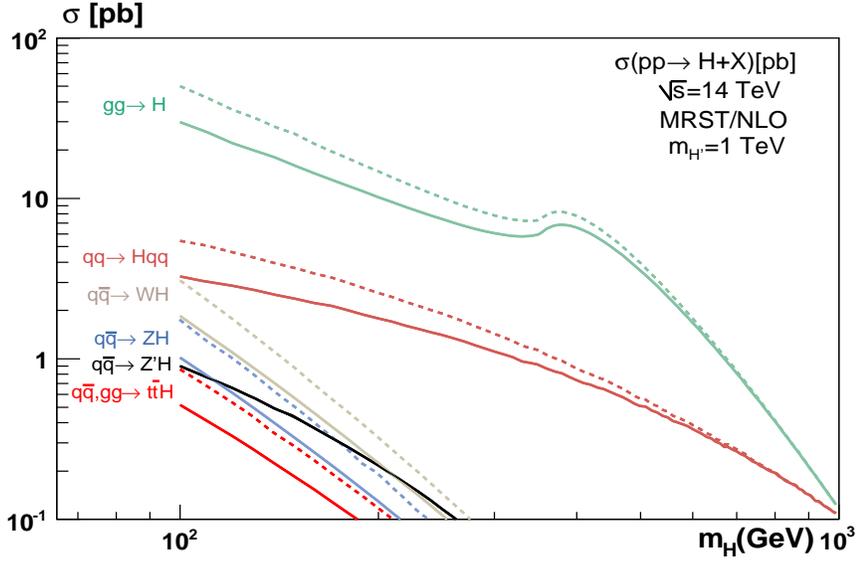, width=13cm, height=8cm,
angle=0}
\end{center}
\par
\vskip-0.5cm \caption{The cross sections of the light Higgs
production as function of $m_H$: $100~ \rm{GeV} \leq m_{H} \leq
1$~TeV, for $m_{H^{\prime}}=1~\rm{TeV}$.}
\label{fig:light-production}
\end{figure}

The cross sections for the Higgs bosons production in these
channels (Equations \ref{ggH}-\ref{qqH}) have been calculated
using the FORTRAN codes: HIGLU, HQQ, V2HV, and VV2HV,
respectively~\cite{Spira}. Extra subroutines have been added to
these programs for the new couplings associated with the two higgs
scalars and the extra gauge boson~\cite{Spira}.
As inputs, we use $v=246$ GeV, $v^{\prime }=1$ TeV, and center of mass energy $%
\surd s=14$ TeV. We also fix the mass of the extra gauge boson at $%
m_{Z^{\prime }}=600$ GeV. The cross sections for the light Higgs boson
production are summarized in Figure~\ref{fig:light-production}. as functions
of the light Higgs mass with $m_{H^{\prime }}=1$ TeV. Figure~\ref%
{fig:heavy-production}, on the other hand, represents the heavy Higgs
productions as functions of $m_{H^{\prime }}$ with $m_{H}=200$ GeV.

As shown in Figure~\ref{fig:light-production}, the salient feature
of this low scale $B-L$ extension is that all cross sections of
the light Higgs production are reduced by about $25-35\%$ in the
interesting mass range: $m_{H}<250$ GeV. As in the SM, the main
contribution to the production cross section comes from the
gluon-gluon fusion mechanism with a few tens of pb. The next
relevant  contribution is given by the Higgs production in the
weak vector boson mechanism, Equation~\ref{qqH}. This contribution
is at
the level of a few pb, as estimated above. Furthermore, the production associated with $%
Z/W$\ is dominant over the production associated with $Z^{\prime }$ for $%
m_{H}<300$ GeV.

\begin{figure}[t]
\begin{center}
\epsfig{file=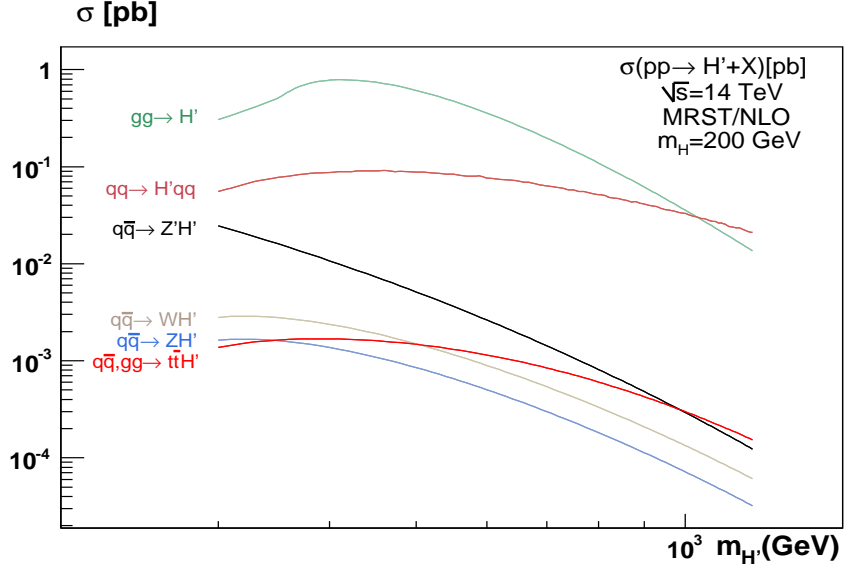, width=13cm, height=8cm,
angle=0}
\end{center}
\par
\vskip-0.5cm \caption{The cross sections of the heavy Higgs
production as function of $m_{H^{\prime}}$: $300~ \rm{GeV} \leq
m_{H^{\prime}} \leq 1$~TeV, for $m_{H}=200~\rm{GeV}$.}
\label{fig:heavy-production}
\end{figure}

Now, we analyze the production of the  heavy Higgs.  It turns out
that its cross sections are smaller than the light Higgs ones. As
shown in Figure~\ref{fig:heavy-production}, all these cross
sections are scaled down by factor ${\cal O}(10^{-2})$, which is
consistent with the result obtained in Equation~\ref{ratio}.
Unlike the light Higgs scenario, the production associated with
$Z^{\prime }$ is dominant over the production associated with Z/W
in agrement with our previous prediction.

\subsection{Higgs Decay}

\begin{figure}[h]
\hspace{3cm}\resizebox{9.5in}{!}{\includegraphics*[0.5in,8.5in][12in,10in]{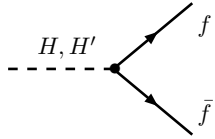}}
\caption{The Feynman diagram for the Higgs boson decays into
fermions.} \label{fig:Fyn3}
\end{figure}

\begin{figure}[h]
\resizebox{9.5in}{!}{\includegraphics*[0.5in,8.25in][12in,10in]{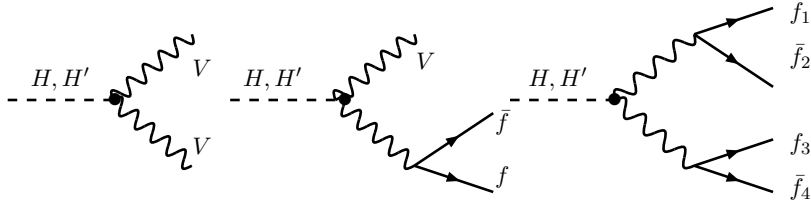}}
\vspace*{-6mm} \caption{Diagrams for the Higgs boson decays into
massive gauge bosons.} \label{fig:Fyn5}
\end{figure}

\begin{figure}[h]
\resizebox{9.5in}{!}{\includegraphics*[0.5in,7.0in][12in,10in]{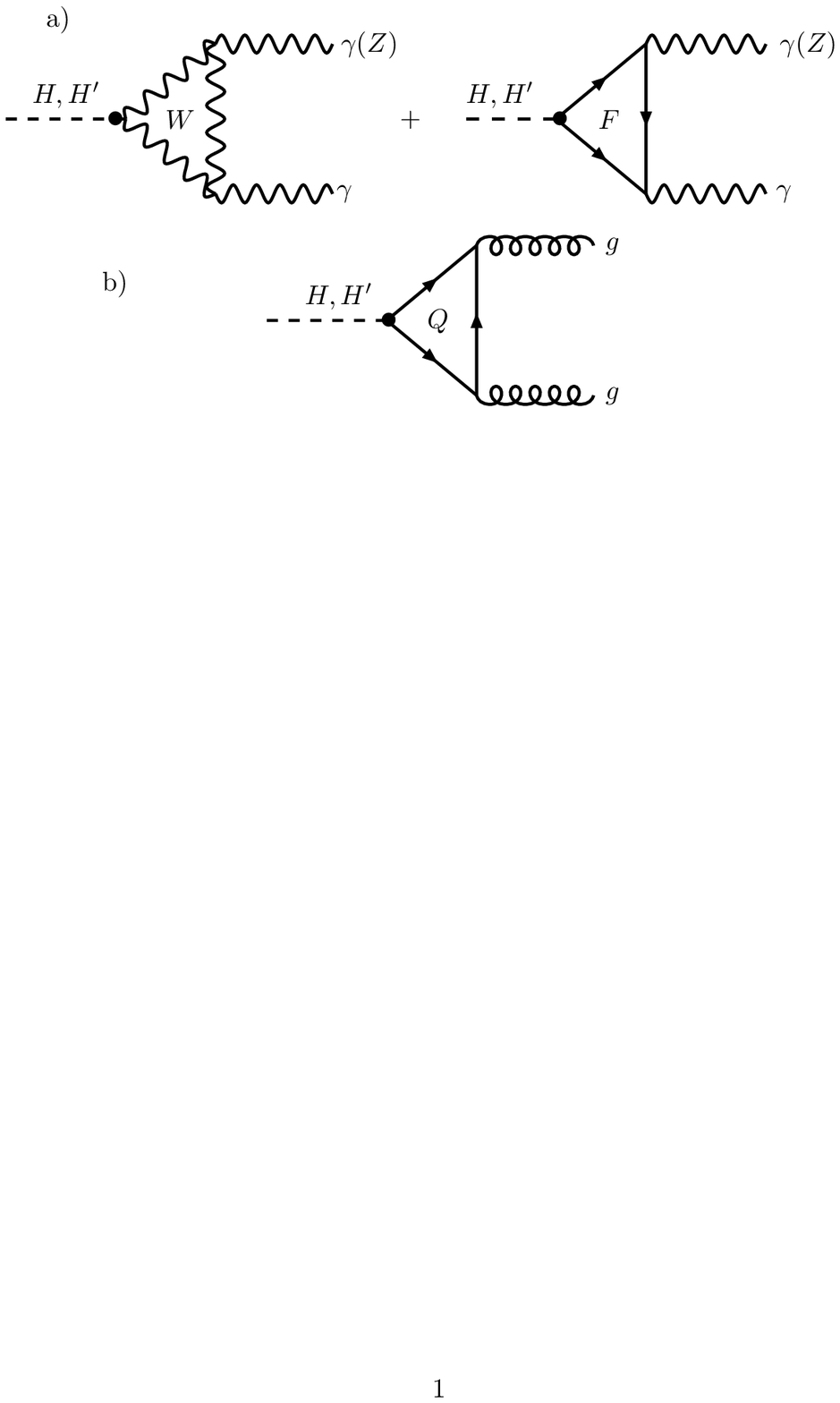}}
\vspace*{-9mm}
\caption{Loop induced Higgs boson decays into a) two photons $(Z\protect%
\gamma$) and b) two gluons.} \label{fig:Fyn6}
\end{figure}

The Higgs particle tends to decay into the heaviest gauge bosons
and fermions allowed by the phase space. The Higgs decay modes can
be classified into three categories: Higgs decays into fermions
(Figure~\ref{fig:Fyn3}), Higgs decays into massive gauge bosons
(Figure~\ref{fig:Fyn5}), and Higgs decays into massless gauge
bosons (Figure~\ref{fig:Fyn6}).


\begin{figure}
\hspace{-4.cm}\resizebox{9.5in}{!}{\includegraphics*[0.5in,8.0in][12in,9.5in]{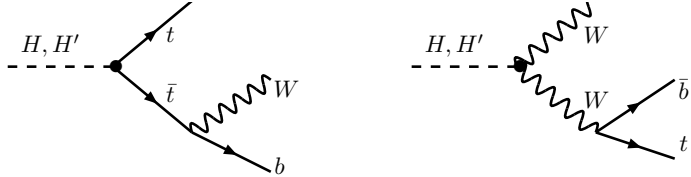}}
\vspace*{-1.4cm} \caption{Diagrams for the three--body decays of
the Higgs boson into $tbW$ final states.} \label{fig:Fyn4}
\end{figure}

The decay widths into fermions are directly proportional to the
$Hff$ couplings
\begin{eqnarray}
\Gamma (H &\longrightarrow &ff)\approx m_{H}\left(\frac{m_{f}}{v}\right)^{2}\left(1-\frac{4m_{f}^{2}}{%
m_{H}^{2}}\right)^{3/2}\cos ^{2}\theta , \\
~\ \ \ \Gamma (H^{\prime } &\longrightarrow &ff)\approx
m_{H^{\prime
}}\left(\frac{m_{f}}{v}\right)^{2}\left(1-\frac{4m_{f}^{2}}{m_{H^{\prime
}}^{2}}\right)^{3/2}\sin ^{2}\theta .
\end{eqnarray}%
In case of the top quark, three-body decays into on-shell and
off-shell states (Figure~\ref{fig:Fyn4}) were taken into
consideration.

\vspace*{1mm}

On the the hand, the decay widths into massive gauge bosons $%
V=Z^{\prime },Z,W$ are directly proportional to the $HVV$
couplings. This includes two-body, three-body, and four-body
decays
\begin{equation}
V\equiv W/Z:~ \Gamma _{H} \approx \frac{m_{H}^{3}}{v^{2}}\left(1-\frac{4m_{V}^{2}}{%
m_{H}^{2}}\right)^{3/2}\!\!\cos ^{2}\theta,~ \ \ \Gamma
_{H^{\prime}}\ \approx
\frac{m^3_{H^{\prime}}}{v^{2}}\left(1-\frac{4m_{V}^{2}}{m_{H^{\prime
}}^{2}}\right)^{3/2}\!\!\sin ^{2}\theta  ,
\end{equation}%
\begin{equation}
~\ \ \ \ V\equiv Z^{\prime }:~ \Gamma _{H}\propto
\frac{m_{H}^{3}}{v^{\prime 2}}\left(1-\frac{4m_{V}^{2}}{%
m_{H}^{2}}\right)^{3/2}\!\!\sin ^{2}\theta ,~\ \ \Gamma
_{H^{\prime }}~\propto  \frac{m^3_{H^{\prime }}}{v^{\prime
2}}\left(1-\frac{4m_{V}^{2}}{m_{H^{\prime
}}^{2}}\right)^{3/2}\!\!\cos ^{2}\theta .
\end{equation}

As shown in Figure \ref{fig:Fyn6}, the massless gauge bosons are
not directly coupled to the Higgs bosons, but they are coupled via
W, charged fermions, and quark loops. This implies that the decay
widths are in turn proportional to the $HVV$ and $Hff$ couplings,
hence they are relatively suppressed.

From the above Equations, one finds that all decay widths of the
light Higgs are proportional to $\cos^2\theta$, except the new
decay mode of $Z^{\prime} Z^{\prime}$. Furthermore, this channel
has a very small contribution to the total decay width. Therefore,
the light Higgs branching ratios (the ratios between the partial
decay widths and the total decay width) have small dependence on
the mixing parameter $\theta$. Thus, it is expected to see no
significant difference between the results of the light Higgs
branching ratios in this model of $B-L$ extension and the SM ones.
On the other hand, the heavy Higgs branching ratios have relevant
dependence on $\theta$.

The decay widths and branching ratios of the Higgs bosons in these
channels have been calculated using the FORTRAN code: HDECAY with
extra subroutines for the new couplings associated with the two
higgs scalars and the extra gauge
boson~\cite{Djouadi:1997yw,Spira}. As in the Higgs production
analysis, we use the following inputs: $v=246$ GeV, $v^{\prime
}=1$ TeV, $m_{Z^{\prime }}=600$ GeV, and c.m. energy $\surd s=14$
TeV.

\begin{figure}[t]
\begin{center}
\epsfig{file=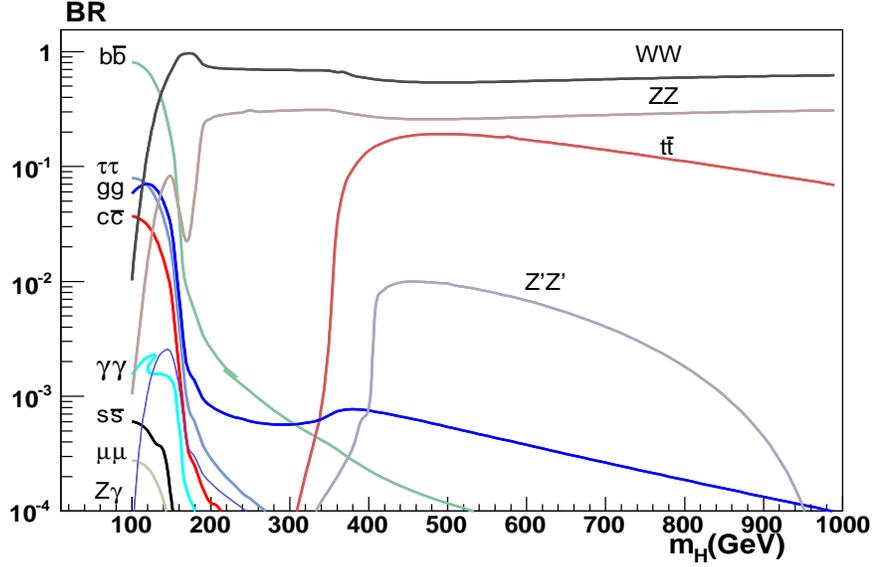, width=13cm, height=8cm, angle=0}
\end{center}
\par
\vskip-0.5cm \caption{The branching ratios of the light Higgs
decay as function of $m_H$ for $m_{H^{\prime}}=1~\rm{TeV}$.}
\label{fig:light-decay}
\end{figure}

\begin{figure}[t]
\begin{center}
\epsfig{file=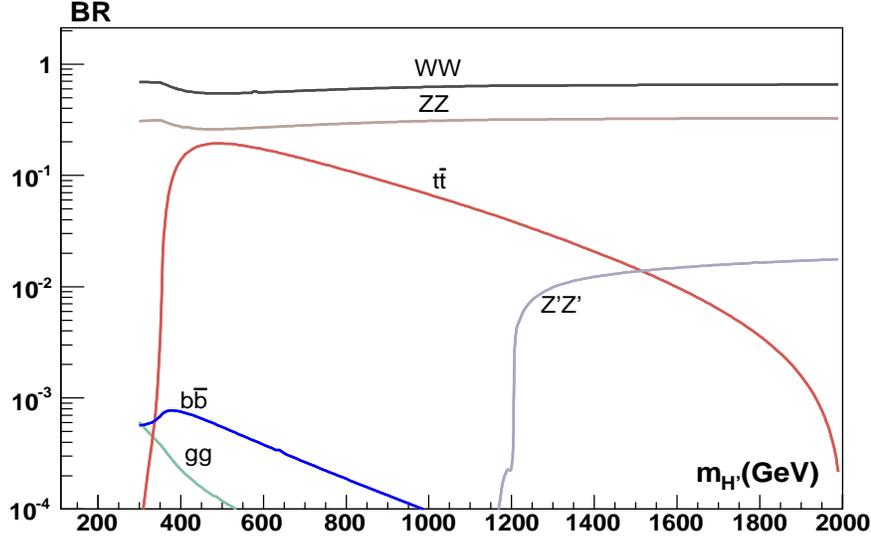, width=13cm, height=8cm, angle=0}
\end{center}
\par
\vskip-0.5cm \caption{The branching ratios of the heavy Higgs
decay as function of $m_{H^{\prime}}$ for $m_{H}=200~\rm{GeV}$}
\label{fig:heavy-decay}
\end{figure}

The decay branching ratios of the light and heavy Higgs bosons are
shown in Figures~\ref{fig:light-decay} and ~\ref{fig:heavy-decay},
respectively, as functions of the Higgs masses. As expected, the
branching ratios of the light Higgs are very close to the SM ones.
In the \textquotedblleft low mass\textquotedblright range: $100$
GeV $<M_{H}<130$ GeV, the main decay mode is $H\rightarrow
b\bar{b}$ with a branching
ratio of $\sim 75-50\%$ . The decays into $\tau ^{+}\tau ^{-}$%
and $c\bar{c}$ pairs come next with branching ratios of order
$\sim 7-5\%$ and $\sim 3-2\%$, respectively. The $\gamma \gamma $
and $Z\gamma $ decays are rare, with very small branching ratios.
In the \textquotedblleft High mass \textquotedblright range:
$m_{H}>130 $ GeV, the $WW$, $ZZ$, and to some extent the
$t\bar{t}$ decays give the dominant contributions. The
$Z^{\prime}Z^{\prime}$ decay arises for quite large Higgs mass
with a small branching ratio $\lsim 1 \%$.

Regarding the heavy Higgs decay branching ratio, one finds that
$H^{\prime} \to WW$ and $ZZ$ are the dominant decay modes, with a
branching ratio of $\sim 70\%$ and $\sim 20\%$, respectively. To a
lower extent, the $t\bar{t}$ and $Z^{\prime}Z^{\prime}$ account
for the remaining branching ratios. Note that these two decay
modes are in particular sensitive to the running mixing angles.
Thus, they have the behaviors shown in
Figure~\ref{fig:heavy-decay}. The other modes give very tiny
contributions and hence they are not shown in this figure.

It is useful to mention that the heavy Higgs may decay to a pair
of the lighter Higgs. The partial decay width of this channel,
which can be expressed by
\begin{equation}
\Gamma (H^{\prime} \longrightarrow HH)\approx
\frac{1}{16\pi\sqrt{2}}
\frac{g^2_{H^{2}H^{\prime}}}{m_{H^{\prime}}}
\left(1-\frac{4m_{H^{\prime}}^{2}}{m_{H}^{2}}\right)^{1/2},
\end{equation}
is suppressed by the tiny $g_{H^{2}H^{\prime}}$ coupling
(Equation~\ref{Hcouplings}) and the relatively large
$m_{H^{\prime}}$. In fact, the resulting branching ratio of this
decay mode is at the level of $10^{-8}$, and hence does not appear
in Figure~\ref{fig:heavy-decay}.

\section{$Z^{\prime}$ decay in $B-L$ extension of the SM}

In this section we study the decay of the extra gauge boson
predicted by the $B-L$ extension of the SM at LHC. In fact, there
are many models which contain extra gauge
bosons~\cite{Hewett:1989rm,Carena:2004xs}. These models can be
classified into two categories depending on whether or not they
arise in a GUT scenario. In some of these models, the $Z^{\prime
}$ and the SM $Z$ are not true mass eigenstates due to mixing.
This mixing induces the couplings between the extra $Z^{\prime }$
boson and the SM fermions. However, there is a stringent
experimental limit on the mixing parameter.
In our model of $B-L$ extension of the SM, there is no tree-level $%
Z-Z^{\prime }$mixing. Nevertheless, the extra $B-L$ $Z^{\prime }$
boson and the SM fermions are coupled through the non-vanishing
$B-L$ quantum numbers.

The interactions of the $Z^{\prime}$ boson with the SM fermions are described by
\be %
{\cal L}^{Z^{\prime}}_{\rm{int}} = \sum_{f}Y^f_{B-L}~ g^{\prime
\prime}~ Z^{\prime}_{\mu}~ f \gamma^{\mu}f. %
\ee %
The decay widths of $Z^{\prime} \to f \bar{f}$ are then given
by~\cite{Carena:2004xs}
\begin{eqnarray}
\Gamma (Z^{\prime } \rightarrow l^+l^-)&\approx& \frac{(g^{\prime
\prime}Y_{B-L}^{l})^{2}}{24\pi }m_{Z^{\prime }}  \nonumber \\
\Gamma (Z^{\prime } \rightarrow q\bar{q})&\approx&\frac{(g^{\prime
\prime}Y_{B-L}^{q})^{2}}{8\pi }m_{Z^{\prime }}\left(1+\frac{\alpha
_{s}}{\pi }\right), ~ q\equiv b,c,s \nonumber \\\Gamma (Z^{\prime
} \rightarrow t\bar{t})&\approx&\frac{(g^{\prime \prime
}Y_{B-L}^{q})^{2}}{8\pi}m_{Z^{\prime}} \left(1-\frac{m_{t}^{2}}{m_{Z^{\prime}}^{2}%
}\right)\left(1-\frac{4m_{t}^{2}} {m_{Z^{\prime
}}^{2}}\right)^{1/2} \nonumber\\
&& \left(1+\frac{\alpha _{s}}{\pi }+O\left(\frac{\alpha
_{s}m_{t}^{2}}{m_{Z^{\prime }}^{2}}\right)\right)
\end{eqnarray}

\begin{figure}[t]
\begin{center}
\epsfig{file=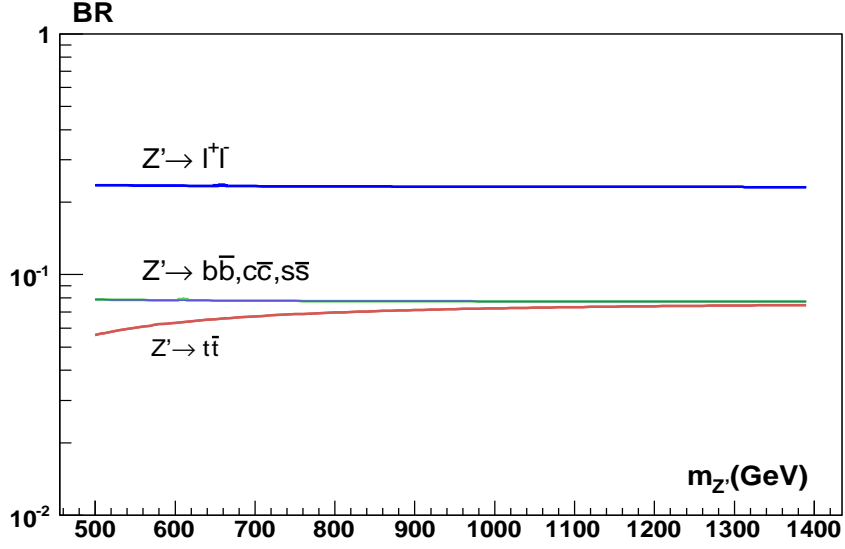, width=13cm,height=8cm,angle=0}
\end{center}
\caption{The decay branching ratios of the extra gauge boson
$Z^{\prime}$ as function of $m_{Z^{\prime}}$.} \label{fig:cdecay}
\end{figure}

Figures~\ref{fig:cdecay} shows the decay branching ratios of
$Z^{\prime }$ as a function of $m_{Z^{\prime }}$. Contrarily to
the SM $Z$ decay, the branching ratios of $Z^{\prime }$
$\rightarrow l^{+}l^{-}$ are relatively high compared to
$Z^{\prime }$ $\rightarrow q\bar{q}$. This is due to the fact that
$\vert Y^l_{B-L} \vert = 3 \vert Y^q_{B-L} \vert$. Thus, one finds
BR($Z^{\prime }$ $\rightarrow l^{+}l^{-})\simeq 20\%$ compared
with BR($Z$ $\rightarrow l^{+}l^{-})\simeq 3\%$. Therefore,
searching for $Z^{\prime }$ can be easily accessible via a clean
dilepton signal, which can be one of the first new physics
signatures to be observed at the LHC.

\section{Light $H^{\prime}$ Scenario}

In this section we discuss the possibility of having
$m_{H^{\prime}} \lsim m_H$ and the phenomenological implications
of this scenario. As shown in section two, the mass of the non-SM
Higgs $m_{H^{\prime}}$ receives a dominant contribution from the
vev of the $B-L$ symmetry breaking $v^{\prime}$ and the self
coupling $\lambda_2$. The $Z^{\prime}$ searches and the neutrino
masses impose a lower limit on $v^{\prime}$: $v^{\prime} \gsim 1$
TeV. The self coupling $\lambda_2$ is essentially unconstrained
parameter. If $\lambda_2 \sim {\cal O}(1)$, then $m_{H^{\prime}}$
is of order TeV as assumed in the previous sections.

There are two other interesting possibilities which have recently
received some attention in the literature. The first one
corresponds to the case of $\lambda_1 v^2 \sim \lambda_2 v^{\prime
2}$, \ie, $\lambda_2 \sim {\cal O}(10^{-2})$. Therefore, one finds
$m_H \sim m_{H^{\prime}}$ and the mixing angle is given by $\theta
\sim \pi/4$. Hence, the two Higgs $H$ and $H^{\prime}$ couple
similarly to the fermion and gauge fields, giving the same
production cross section and decay branching ratio. Therefore, the
distinguish between $H$ and $H^{\prime}$ at LHC in this type of
models is rather difficult. This scenario is usually known as {\it
intense Higgs coupling}~\cite{Boos:2002ze}.

The second possibility concerns the case of $\lambda_2 \lsim
10^{-3}$, in which one obtains $m_{H^{\prime}} \ll m_H$. In fact,
LEP and Tevatron direct searches do not exclude a light Higgs
boson with a mass below $60$ GeV. Such light Higgs may have
escaped experimental detection due to the suppression of its cross
sections. Therefore, a window with a very light Higgs mass still
exist.

Having $\lambda_2 \lsim 10^{-3}$ implies that $\lambda_3$ is also
less than $10^{-3}$. In this respect, the Higgs masses are
approximately given by %
\bea %
m_H &\simeq& \sqrt{\lambda_1} v, \\
m_{H^{\prime}} &\simeq& {\cal O}\left(\frac{\lambda_3
v^{\prime}}{\lambda_1 v}\right) \simeq {\cal O}(10^{-2}) \rm{GeV},
\eea %
and the coupling $g_{HH^{\prime} H^{\prime}}$ in Equation
\ref{Hcouplings} becomes very small. Thus, the decay $H \to
H^{\prime}H^{\prime}$ is not comparable to the decay into other SM
particles. The phenomenology of this scenario, derived from
different SM extensions, has been studied in
details~\cite{O'Connell:2006wi},\cite{Krawczyk:2000kf}. In
addition, this light scalar particle has been considered as an
interesting candidate for dark matter \cite{Boehm:2006gu}.

\section{ Conclusions}

In this paper we have considered the TeV scale $B-L$ extension of
the SM. We provided a comprehensive analysis for the phenomenology
of the SM like Higgs, the extra Higgs scalar, and the extra gauge
boson predicted in this model, with special emphasize on their
potential discovery at the LHC.

We have shown that the cross sections of the SM-like Higgs
production are reduced by $\sim 20\%-30\%$ in the mass range of
$\sim 120 -250$ GeV compared to the SM results. On the other hand,
the implications of the $B-L$ extension to the SM do not change
the decay branching ratios. Moreover, we found that the extra
Higgs has relatively small cross sections, but it is accessible at
LHC. Finally, we showed that the branching ratios of $Z^{\prime}
\to l^+ l^-$ are of order $\sim 20\%$ compared to $\sim 3\%$ of
the SM $BR(Z\to l^+ l^-)$. Hence, searching for $Z^{\prime}$ is
accessible via a clean dilepton signal at LHC.

\section*{Acknowledgment}
This work is partially supported by the ICTP under the
OEA-project-30.

\end{document}